\documentstyle[sprocl]{article}
\input epsf

\begin{document}

\title{Chaos and High--Energy Collisions}

\author{S. G. Matinyan}

\address{Yerevan Physics Institute, Armenia \\
and \\
Department of Physics, Duke University, Durham, NC 27708-0305}

\maketitle\abstracts{The investigations of the gauge boson wave packets 
collisions in SU(2) Higgs model are presented.  The evolution of the gauge 
and Higgs fields is studied as a function of the amplitude and the mass 
ratio of $M_H/M_W$.  We visualize the restoration of spontaneously broken 
gauge symmetry during collisions.}

\section{Introduction. Infrared Instabilities and Dynamical Chaos}

Recently, the nonperturbative treatment of the high multiplicity
scattering amplitudes which are not suppressed in the weak coupling
limit has attracted attention \cite{RT92,GNV92,Gong94,Hu95}.  It is
quite natural in the connection with the exciting expectations that
the rate of the baryon-number violating electroweak processes,
non-perturbative in essence \cite{Hooft76}, might be significant at
ultra-high energies \cite{Ring90}.  The semiclassical technique,
however, meets with difficulties in the study of $2 \to$ many
particles amplitude, since the {\it initial} state of few highly
energetic particles is not semiclassical at all (see, e.g.
\cite{Vol95}).

In the extreme non-perturbative classical treatment of the high energy
multiparticle amplitude, the question is the following:  Does there
exist a mechanism for energy transfer from high frequency modes,
corresponding to two (or few) initial high energy particles, to low
frequency modes representing a multiparticle final states?
At first glance, the answer to this question, formulated in terms of
nonlinear dynamics, seems to be affirmative since the gauge field
equations are nonlinear.  However, the studies of $(1+1)$-dimensional
Abelian Higgs model \cite{RT92} and $\lambda\varphi^4$-theory
\cite{GNV92} have shown no indication for a mechanism providing the
coupling between the initial high and the final low frequency modes.

Of course, the gauge field nonlinearities inherent in the non-Abelian
gauge theories and which are absent in the Abelian
models,\footnote{The nonlinearities of Abelian models due to the
Higgs-gauge fields and the Higgs self-coupling, as we see below, are
not important at the high energy gauge boson collisions.} 
are essential and, in general, lead to the infrared instabilities.

One may say that for non-Abelian gauge theories the infrared
instabilities are not an exception, but rather a rule, and they are
intimately connected with i) the masslessness of the gauge field, ii)
its isospin (color) charge, and iii) the gyromagnetic ratio of the
gluon equal 2.  Nonlinearity itself is not enough to furnish the above 
mentioned coupling between fast and slow modes, as one can argue from 
negative results of \cite{RT92,GNV92}.

From a more general point of view, the observed inability of the
nonlinearity alone to provide a mechanism for the formation of the
inelastic final states is intimately connected with the integrable
nature of the classical systems considered in \cite{RT92,GNV92}.  It
is well known that non-Abelian gauge theories are nonintegrable in the
classical limit and exhibit dynamical chaos \cite{Mat81,Mul92} (see
also \cite{Biro94} for details and extended literature).

This dynamical stochasticity of the non-Abelian gauge fields together
with their mentioned instability are two possible sources of the
mechanism for the coupling between high and low frequency
modes.\footnote{It is not excluded that these two sources have a
common deep origin, though I am unable to prove this assertion on
theorem-like grounds.} At the same time, it is not superfluous to
recall the role of the Higgs condensate in the suppressing of the
chaos of the non-Abelian gauge fields \cite{MST81}.

In \cite{Hu95} we studied the collision of two SU(2) gauge field wave
packets.  As we expected, based on our previous results \cite{Gong94},
the collisions of essentially non-Abelian initial configurations
trigger the decay of initial states into many low frequency modes with
dramatically different momentum distributions, whereas for Abelian
configurations wave packets pass through each other without
interaction.

Here I will present the study of the collisions of wave packets in the
SU(2) Higgs model where the fundamental excitations of the gauge-field
are massive.

\section{Collisions of Classical Wave Packets in SU(2) Higgs Model}

\subsection{SU(2) Higgs Model}

We briefly describe the spontaneously broken SU(2) model with an
isodoublet Higgs field $\Phi$.  This model retains the most relevant
ingredients of the electroweak theory.  The action of this model is
given by
\begin{eqnarray}
S &= &\int d^3xdt \left\{ -{1\over 2}{\rm tr}\left(
F_{\mu\nu}F^{\mu\nu}\right) + {1\over 2}{\rm tr}\left[({\cal D}_{\mu}
\Phi)^{\dagger}{\cal D}^{\mu}\Phi\right] \right . \nonumber \\
&&\left . \quad - \lambda \left[{1\over 2}{\rm tr} (\Phi^+\Phi) -v^2\right]^2
\right\}. \label{e2}
\end{eqnarray}
with ${\cal D}_{\mu} = \partial_{\mu}-ig A_{\mu}^a\tau^a/2$, 
$F_{\mu\nu}\equiv F_{\mu\nu}^a{\tau^a\over 2} = {i\over g}[{\cal
D}_{\mu}, {\cal D}_{\nu}]$ and $\Phi = \phi^0 - i\tau^a\phi^a$;
$\tau^a(a=1,2,3)$ are Pauli matrices.  $v$ is a vacuum expectation
value $(v.e.v.)$ of the neutral component of the scalar field.

By proper scaling transformations of the space-time coordinates and
fields, it is easy to see that the action (\ref{e2}) and the
corresponding equations of motion possesses only a single parameter
\begin{equation}
{\lambda\over g^2} = {M_H^2\over 8M_W^2} 
\end{equation}
($M_H = 2v\sqrt{\lambda}\; {\rm and}\; M_W = {gv\over\sqrt{2}}$
are the tree masses of Higgs and gauge $W$-bosons respectively; $v =
174$ GeV.)  However, in the simulation of the wave packet collisions,
initial conditions introduce extra physical parameters.

We work in the unitary gauge where only physical excitations appear:
\begin{eqnarray}
\Phi &= &\left( v+ {\rho\over\sqrt{2}}\right) U(\theta) \nonumber \\
A_{\mu} &= &U(\theta)W_{\mu} U^{-1}(\theta) + {i\over g}
\left( \partial_{\mu}U(\theta)\right) U^{-1}(\theta) \label{e3}
\end{eqnarray}
with $U(\theta) = \exp (i\tau^a\theta^a)$.

The real field $\rho$ describes the oscillations of the scalar field
about its $v.e.v$., and $W_{\mu}$ is the $W$-boson field:
\begin{equation}
(\partial_{\mu}\partial^{\mu}+M_H^2) \rho + 3\sqrt{2} \lambda v\rho^2
+ \lambda\rho^3 - {1\over 4} g^2W_{\mu}^aW^{a\mu}\rho^2 - {1\over
2\sqrt{2}} g^2vW_{\mu}^a W^{a\mu} = 0, \label{e4}
\end{equation}
\begin{equation}
[{\cal D}_{\mu},F^{\mu\nu}] + \left( M_W^2 + {1\over\sqrt{2}} g^2v\rho
+ {1\over 4}g^2\rho^2\right) W^{\nu} = 0. \label{e5}
\end{equation}
We emphasize that the gauge field acts as a source for Higgs
excitations in (\ref{e4}) (last term in (\ref{e4})).  This permits us
to consider the $W$-field classically. 

\subsection{Scattering of Wave Packets}

Our numerical study is based on the Hamiltonian formulation of lattice
SU(2) gauge theory \cite{KS75} (see \cite{Hu95,GongT} for details).
We work on an one-dimensional lattice with a size $L=Na$ ($N$ is the
number of lattice sites, $a$ is the lattice spacing).

To implement the temporal gauge $W_0^a=0$, most convenient in the
Hamiltonian formulation of the lattice gauge theory, we ``collide''
transverse $W$-bosons, for which the relation
$\partial_{\mu}W^{a\mu}=0$ holds.

The initial configuration is given by two well-separated right- and
left- moving Gaussian wave packets originally centered at $z_{R(L)}$
with average momenta ${\bf k}=(0,0,\bar k)$ and width $\Delta k\; 
(\Delta k\ll \bar k)$:
\begin{eqnarray}
W^{c,\mu} &= &W_R^{c,\mu} + W_L^{c,\mu} \nonumber \\
W_{R(L)}^{c,\mu} &= &\delta^{\mu2} n_{R(L)}^c \psi(z-z_{R(L)},\pm t)
\label{e6}
\end{eqnarray}
with $n_{R(L)}^c$ the unit isospin vectors.  To specify the profile
function $\psi(z,t)$ we take, for a right- moving wave packet centered
at $z=0$ at $t=0$,
\begin{equation}
\psi(z,t) = {1\over \sqrt{\pi^{3/2}\bar\omega\Delta k\sigma}}
\int_{-\infty}^{\infty} dk\; e^{-(k-\bar k)^2/2(\Delta k)^2}
\cos(\omega t-kz) \label{e7}
\end{equation}
with $\omega = (k^2+M_W^2)^{1/2}$, and the normalization is fixed by
requiring energy equal to $\bar\omega$ per cross-sectional area
$\sigma$.

From (\ref{e7}) we get the initial conditions for $\psi(z,0)$ and
${d\psi\over dt}(z,t)\big\vert_{t=0}$ which we don't give explicitly here.  
The initial condition for the Higgs field is given by the vacuum solution 
at $t=0$: $\phi^0=v,\; \phi^a=0,\; \dot\phi^0 = \dot\phi^a=0$.  $\left(. 
\equiv {d\over dt}\right)$.  It is possible to see that the initial
conditions introduce three new dimensionless parameters $\bar k/v,
\;{\Delta k\over v}$ and ${\sigma v^2\over g^2}$ in addition to
${\lambda\over g^2}$ (or $M_H/M_W$) (we fix in the following $g=0.65$).

One more parameter appears in the initial conditions---the angle
$\theta_c$ between the relative orientation of isospins of two wave
packets.  Non-zero $\theta_c$ corresponds to the essentially
non-Abelian configuration, $\theta_c=0$ (parallel isospins) gives the
initial pure Abelian configurations.

For the pure Yang-Mills wave packet collisions, the non-linearity is
due to the self-interaction of gauge fields.  As it was established in
\cite{Hu95} for $\theta_c=0$ no indications of the final inelastic
states had a place:  wave packets passed through each other without
interaction.  On the contrary, for $\theta_c\not= 0$ collisions
resulted in strongly inelastic final states \cite{Hu95}.  It is
remarkable that the inelastic patterns remain qualitatively similar
for $\theta_c$ as small as $\sim 10^{-12}$, clearly connecting these
phenomena with the dynamical chaos of the non-Abelian gauge fields.

For the Yang-Mills-Higgs system, the situation is more involved due to
the additional non-linearities induced by the gauge field-Higgs and
the Higgs self-couplings.

Figures 1 and 2 (top rows) show a few ``snapshots'' of the space-time
development of the colliding $W$-boson wave packets for parallel (Fig.
1) and orthogonal (Fig. 2) isospin orientations.  The figures show the
absolute value of the scaled gauge field amplitude $\vert A\vert/v$.  
For parallel isospin orientations, the result of the
``collisions'' is a slight distortion of the initial wave packets
showing no sign of the inelasticity.  Decreasing of the energy 
$(\bar k = \pi/25)$ shows a small inelasticity for $\theta_c = 
0$.\footnote{This fact is easily explained by the consideration of the 
tree diagrams in $WW$-scattering since the scalar exchanges are 
decreased with energy.}  As is seen from the top row 
of Fig. 2, for $\theta_c=\pi/2$, final states are strongly inelastic.

\begin{figure}
\def\epsfsize#1#2{.6#1}
\centerline{\epsfbox{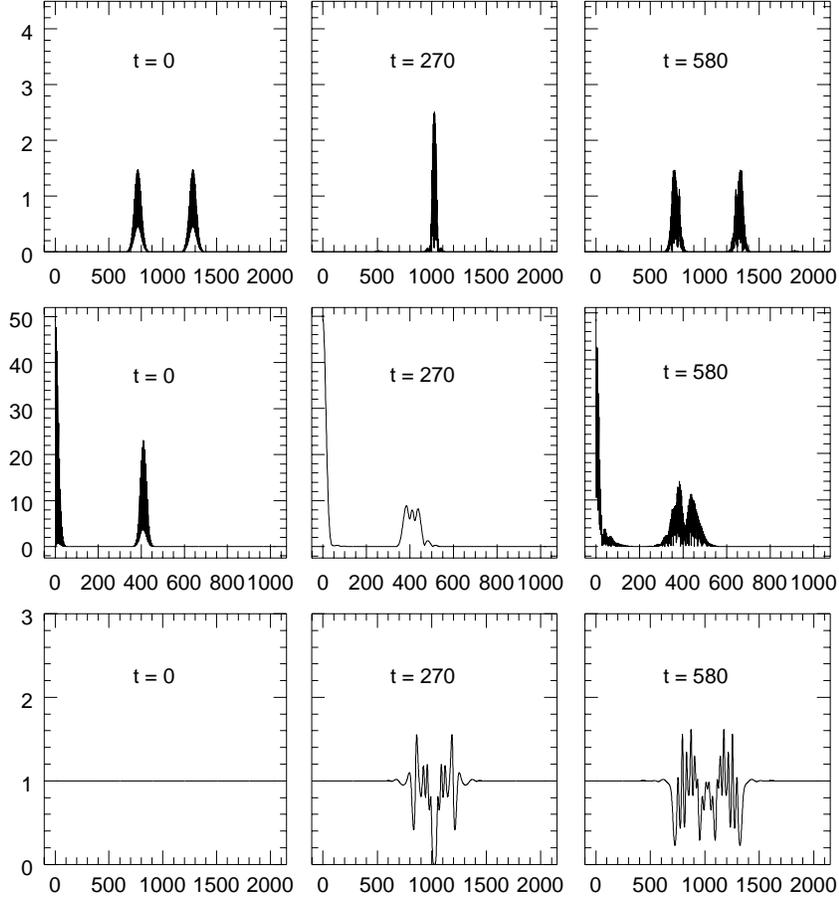}}
\caption{Collision of two $W$-wave packets with parallel isospin
polarizations.  We choose $M_H=M_W=0.126$, $\bar k = \pi/5$, $\Delta k
= \pi/100$, $g=0.65$, and $\sigma = 0.336$.  This simulation, as well
as all others below, was performed on a lattice of length $L=2048$ and
lattice spacing $a=1$.  The top row shows the space-time evolution of
the scaled gauge field amplitude $\vert A\vert/v$, the median row
exhibits the corresponding Fourier spectra of the gauge field energy
density, and the bottom row shows the space-time evolution of the
scaled Higgs field $\vert\Phi\vert^2/v^2$.  The abscissae of top and
bottom rows are labelled in units of the lattice spacing, and the
abscissa of the median row is in units of $\sigma/1024$.}
\end{figure}

\begin{figure}
\def\epsfsize#1#2{.6#1}
\centerline{\epsfbox{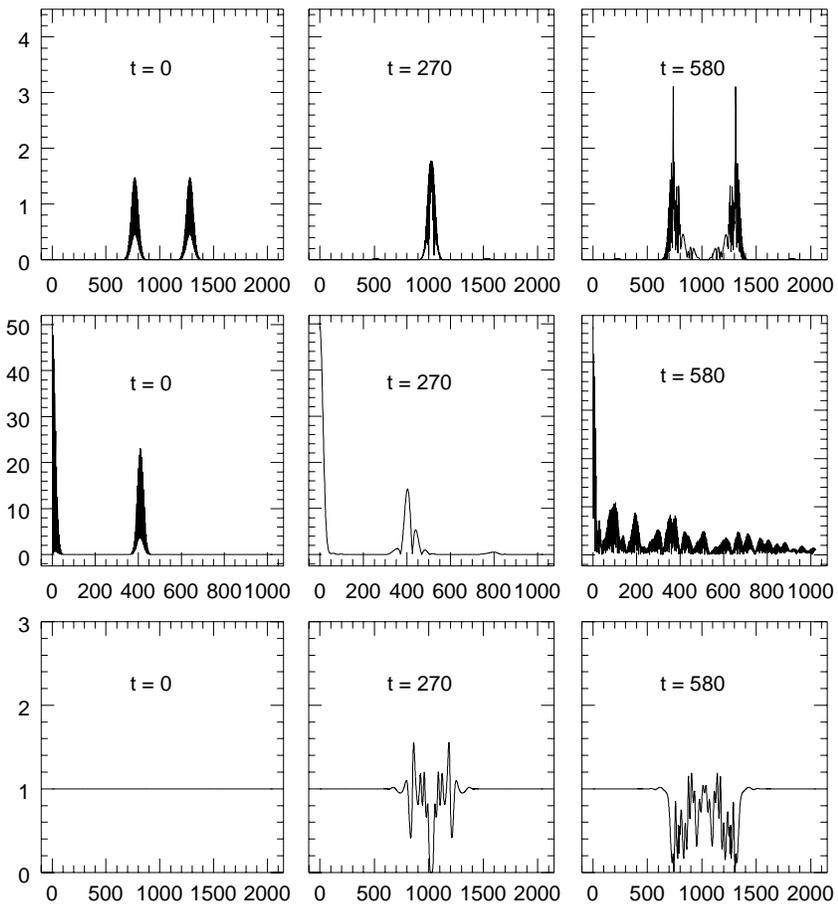}}
\caption{Same as for Figure 1, for orthogonal polarizations.}
\end{figure}

The difference between the two cases ($\theta_c=0,\;\pi/2)$ is more
striking by looking at the evolution of the absolute value of the
Fourier transform of the gauge-invariant energy density (scaled by
$v^2$) (median row in Figs. 1 and 2).

The bottom rows of Figs. 1 and 2 illustrate the time evolution of the
Higgs field excitations around its $v.e.v.\; v$ (scaled to unity).
Here we have plotted the $\vert\Phi\vert^2/v^2$ as a function of space
coordinate at three different times.

\section{Symmetry Restoration in WW collisions}

As is seen from Figures 1-2 (bottom rows), for not very large
$r=M_H/M_W$ the Higgs field oscillates not about its $v.e.v. \;v$ but
rather about zero.  This suggests that the collisions of the gauge
boson wave packets, accompanied by energy transfer from gauge field to
Higgs field, lead to the restoration of the broken SU(2) symmetry.
This pehnomenon occurs for the large gauge fields amplitude (see
\cite{KL72,KP76}).  Indeed, (\ref{e4}), describing the excitations of
the scalar field about the Higgs vacuum $\vert\Phi\vert = v$, has
another exact solution $\rho=-\sqrt{2}v$ $(\vert\Phi\vert = 0$, see
(\ref{e3})) with $W_{\mu}^c$ being arbitrary.  In terms of the small
excitations $\chi=\rho+v/\sqrt{2}$ about $\Phi=0$, for (\ref{e4}) and
(\ref{e5}) we have:
\begin{eqnarray}
\left[ \partial_{\mu}\partial^{\mu} - {M_H^2\over 2} \left( 1 +
{g^2W^2\over 8\lambda v^2}\right)\right] \chi + \lambda\chi^3 &= &0
\label{e10} \\
\left[ {\cal D}_{\mu},F^{\mu\nu}\right] + {1\over 4} g^2\chi^2W^{\nu}
&= &0 \label{e11}
\end{eqnarray}
where $W^2(x) \equiv W_{\mu}^a(x) W^{a\mu}(x) = -(W_i^a(x))^2 <0$ for
our choice of the transverse $W$-wave packets.  $W^2(x)$ is always
negative for time-like bosons.\footnote{We recall that the luminosity
of transverse $W$-bosons generated by the energetic fermions is much
higher than that of longitudinal ones and increases with energy
\cite{CG84}.}

Eq. (\ref{e11}) describes the massless $W$-boson.  From (\ref{e10}) and
(\ref{e11}), we have the effective potential for $\chi$-excitations:
\begin{equation}
V(\chi,W^2) = -\lambda v^2(1-\eta) \chi^2 + {\lambda\over 4}\chi^4,
\label{e12}
\end{equation}
where we introduce $\eta = {g(W_i^a)^2\over 8\lambda v^2} = {1\over
r^2} \langle W_i^{a^2}\rangle$ as a parameter where the intensity
$(W_i^a)^2$ of the high frequency gauge pulses is replaced by its
space-time average $\langle W^2\rangle$.  Depending on whether $\eta
>1$ or $\eta<1$, the potential (\ref{e12}) has two different {\it
stable} minima:
\begin{eqnarray}
\eta < 1:\quad \chi_{\rm min} &= &\pm \sqrt{2} v(1-\eta)^{1/2}, \quad
{\rm i.e.}\; \vert\phi\vert = v(1-\eta)^{1/2}, \label{e13} \\
\eta > 1: \quad \chi_{\rm min} &= &0, \quad {\rm i.e.}\; \Phi = 0.
\label{e14}
\end{eqnarray}
Stable excitations about these ``vacua'' have the following squared
masses:
\begin{eqnarray}
\tilde M_W^2 &= &M_W^2(1-\eta) \theta (1-\eta) \label{e15} \\
\tilde M_W^2 &= &{M_H^2\over 2} \vert 1-\eta\vert [1+\theta(1-\eta)].
\label{e16}
\end{eqnarray}

Thus for $\eta>1$, the symmetry is restored and the scalar
oscillations occur about the symmetrical state $\Phi=0$ with the zero
effective mass of the $W$-boson.

For $\eta<1$, the vacuum is changed gradually as $(1-\eta)^{1/2}$.
For this case, $\tilde r = \tilde M_H/\tilde M_W=r$.  In Figure 3,
first column, the space development of the $WW$ collision is shown for
$\eta=1.32$ for different times.  As seen from this figure, at time
$t\approx 300$ wave packets collide and then begin to separate.  Just
about at this time one expects to observe the restoration of symmetry,
i.e. the oscillations of the scalar field $\phi$ about a new ground
state located below the ``old'' vacuum $\vert\phi\vert=v$.  After the
separation of the wave packets $(t>300)$ the scalar field excitations
tend again to oscillate about ``old'' vacuum, i.e. the gauge symmetry
is broken again.  The second column exhibits the space-time evolution
of the $\vert\phi\vert/v$.  The third column shows the Higgs field
smoothed over 50 lattice sites in order to facilitate a comparison
with the definition (\ref{e12}) of the parameter $\eta$ in terms of
the averaged strength of the $W$-boson field.
\bigskip

\begin{figure}
\def\epsfsize#1#2{.6#1}
\centerline{\epsfbox{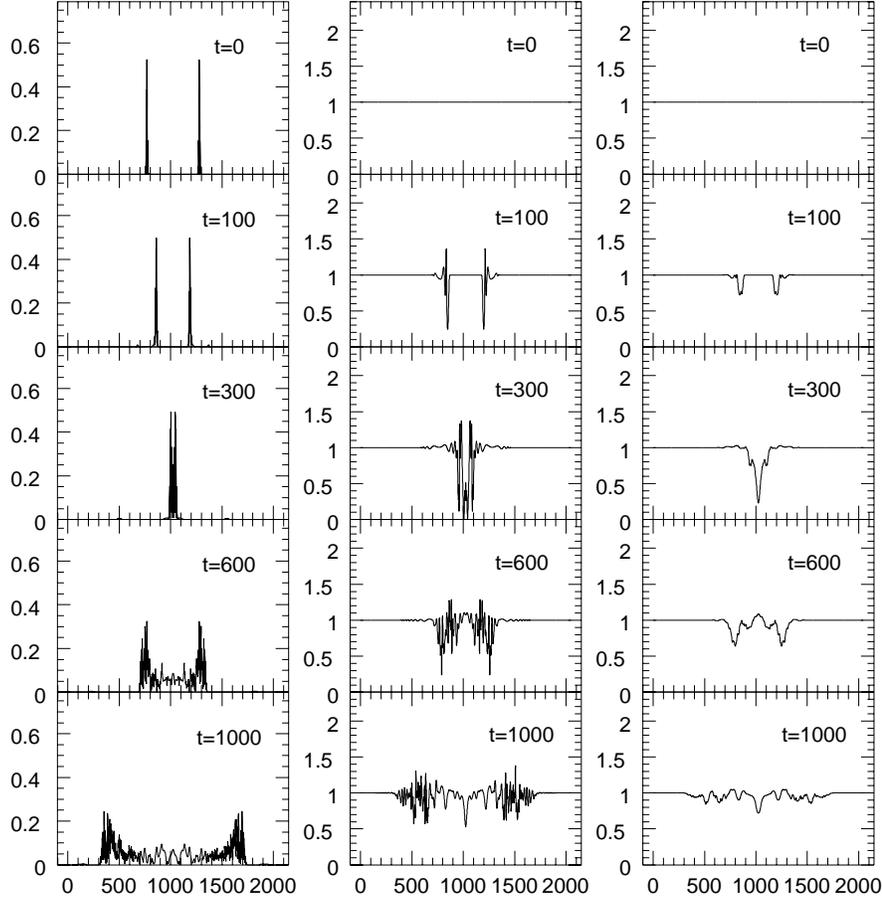}}
\caption{Space-time development of symmetry restoration induced by two
colliding gauge wave packets with orthogonal isospins in the presence
of the Higgs vacuum condensate.  First column shows the scaled gauge
field $\vert A\vert/v$ as a function of space coordinate $z$ five
chosen times.  Second column demonstrates the corresponding space-time
evolution of the scaled Higgs field $\vert\Phi\vert/v$.  Third column
shows the scaled Higgs field after smoothing over 50 lattice sites.
This simulation was done on a lattice of sites $n=2048$ and lattice
spacing $a=1$.  The parameters were $\bar k=\pi/4$, $\Delta k=\pi/16$,
$M_H=M_W=0.15$, $g=0.65$, and $\sigma =1$.}
\end{figure}

\section{Concluding Remarks}

In the numerical studies of the collisions classical wave packets of
transversely polarized gauge bosons with non-parallel isospin
orientations in the broken SU(2) gauge theory we have found evidence 
for the creation of final states with strongly ``inelastic'' events for 
a wide range of the essential parameters of the problem.  

We have observed and visualized the process of the SU(2) symmetry 
res-toration in some finite space-time region as a result of the 
collisions of the intense gauge pulses.  At last but not least, it is 
important to emphasize the observed correlation between the occurence 
of the inelastic events (for non-parallel isospin configurations) and 
the restoration of the symmetry in high energy collisions.  Both these 
phenomena require the same order of the amplitude of the initial 
configurations.
\vfill
\eject

\subsection*{Acknowledgements}

It is my great pleasure to thank C. R. Hu, B. M\"uller and D. Sweet
for enlightening, stimulating discussions and collaboration.  I am
grateful also to K. Rajagopal and R. Singleton for useful comments.
This work was supported in part by grant DE-FG02-96ER40945 from the
U.S. Department of Energy, and by the North Carolina Supercomputing
Center.

\end{document}